\RequirePackage{fix-cm}
\documentclass[smallextended]{svjour3}       % onecolumn (second format)
\usepackage{natbib}

\smartqed  % flush right qed marks, e.g. at end of proof
\usepackage{graphicx}
\journalname{Empirical Software Engineering}
\usepackage{amsmath,amssymb,amsfonts}
\usepackage{algorithmic}
\usepackage{textcomp}
\usepackage{rotating}
\usepackage{multirow}
\usepackage{booktabs}
\usepackage{tabularx}
\usepackage{mdframed}
\usepackage{longtable}

\usepackage{listings}
\usepackage{xcolor}
\usepackage{framed}
\usepackage{caption}
\usepackage{subcaption}

\newcommand\YAMLcolonstyle{\color{red}\mdseries}
\newcommand\YAMLkeystyle{\color{black}\bfseries}
\newcommand\YAMLvaluestyle{\color{blue}\mdseries}

\makeatletter

\newcommand\language@yaml{yaml}

\expandafter\expandafter\expandafter\lstdefinelanguage
\expandafter{\language@yaml}
{
  captionpos=b,
  frame=single,
  keywords={true,false,null,y,n},
  keywordstyle=\color{darkgray}\bfseries,
  basicstyle=\YAMLkeystyle,                                 % assuming a key comes first
  sensitive=false,
  comment=[l]{\#},
  morecomment=[s]{/*}{*/},
  commentstyle=\color{purple}\ttfamily,
  stringstyle=\YAMLvaluestyle\ttfamily,
  moredelim=[l][\color{orange}]{\&},
  moredelim=[l][\color{magenta}]{*},
  moredelim=**[il][\YAMLcolonstyle{:}\YAMLvaluestyle]{:},   % switch to value style at :
  morestring=[b]',
  morestring=[b]",
  literate =    {---}{{\ProcessThreeDashes}}3
                {>}{{\textcolor{red}\textgreater}}1     
                {|}{{\textcolor{red}\textbar}}1 
                {\ -\ }{{\mdseries\ -\ }}3,
}

\usepackage{hyperref}

\begin{document}

\newcommand{\atoml}{\texttt{atoml}}

\newcommand\setrow[1]{\gdef\rowmac{#1}#1\ignorespaces}
\newcommand\clearrow{\global\let\rowmac\relax}
\clearrow

\title{Smoke Testing for Machine Learning: Simple Tests to Discover Severe Bugs}
\titlerunning{Smoke Testing for Machine Learning}

\author{
Steffen Herbold \and
Tobias Haar}

\institute{
Steffen Herbold\\Institute of Software and Systems Engineering, TU Clausthal, Germany\\
\email{steffen.herbold@tu-clausthal.de}
\vspace{5pt}\\
Tobias Haar\\Institute of Computer Science, University of Goettingen, Germany\\
\email{tobias.haar@stud.uni-goettingen.de}
}

\date{Received: date / Accepted: date}

\maketitle

\begin{abstract}
Machine learning is nowadays a standard technique for data analysis within software applications. Software engineers need quality assurance techniques that are suitable for these new kinds of systems. Within this article, we discuss the question whether standard software testing techniques that have been part of textbooks since decades are also useful for the testing of machine learning software. Concretely, we try to determine generic and simple smoke tests that can be used to assert that basic functions can be executed without crashing. We found that we can derive such tests using techniques similar to equivalence classes and boundary value analysis. Moreover, we found that these concepts can also be applied to hyperparameters, to further improve the quality of the smoke tests. Even though our approach is almost trivial, we were able to find bugs in all three machine learning libraries that we tested and severe bugs in two of the three libraries. This demonstrates that common software testing techniques are still valid in the age of machine learning and that considerations how they can be adapted to this new context can help to find and prevent severe bugs, even in mature machine learning libraries. 

\end{abstract}

\section{Introduction}
\label{sec:introduction}

Machine learning is nowadays a standard technique for the analysis of data in many research and business domains, e.g., for text classification~\citep{Collobert2008}, object recognition~\citep{Donahue2014}, credit scoring~\citep{Huang2007}, online marketing~\citep{Glance2005}, or news feed management in social networks~\citep{Paek2010}. Different machine learning libraries were developed throughout the years, e.g., Weka~\citep{WEKA, Frank2016}, SparkML~\citep{Zaharia2010}\footnote{\url{https://spark.apache.org/docs/latest/ml-guide.html}}, scikit-learn~\citep{Pedregosa2011}, caret~\citep{Kuhn2018}, and TensorFlow~\citep{Abadi2016}. Machine learning research focuses mostly on the development of new and better techniques for machine learning and the adoption of machine learning techniques in new domains. When people speak about testing machine learning algorithms, they usually refer to the evaluation of the performance that the algorithm achieves for a given problem, e.g., in terms of \textit{accuracy}, \textit{precision}, or \textit{recall}. Software engineering researchers are among these domain scientists that use machine learning to solve software engineering problems. 

The quality assurance of machine learning algorithms recently gained traction as a novel research topic~\citep{Zhang2020, Braiek2020}. Within this article, we focus on the testing of training and prediction algorithms. The testing of such algorithms is challenging, especially due to the lack of an obvious test oracle, i.e., an entity that can determine correct predictions. Parts of the literature even state that the only possible test oracle are users~\citep{Groce2014}. The current literature focuses on solving specific issues of algorithms or problems in a domain application, e.g., through metamorphic testing as a replacement for the test oracle (e.g., \citep{Murphy2008, Xie2011, Zhang2011, Nakajima2016, Pei2017, Ding2017, Tian2018}), i.e., testing software by modifying the inputs and using the prior results as test oracle for the outcome of the algorithm on the modified input. The drawback of this often relatively narrow focus of the literature on specific algorithms (e.g., a certain type of neural network) or even use cases is that, while the solutions often yield good results for this case, it can be difficult to transfer the results to other contexts. Moreover, the solutions are often highly technical to achieve the maximum benefit for a specific problem and not generalizable to machine learning algorithms in general. 

Within this article, we take a step backwards. Instead of focusing on optimal solutions for specific algorithms or use cases, we look at machine learning in general and try to determine if we can derive broadly applicable best practices from textbook software testing techniques. This research is driven by the following research question. 

\begin{itemize}
    \item[\textbf{RQ:}] What are simple and generic software tests that are capable of finding bugs and improving the quality of machine learning algorithms?
\end{itemize}

Such tests could become part of a standard toolbox of best practices for developers of machine learning algorithms. We use the notion of \textit{smoke testing} to drive our work. Smoke tests assert ``that the most crucial functions of a program work, but not bothering with finer details''.\footnote{\url{https://glossary.istqb.org/search/smoke\%20test}} The most crucial function of machine learning algorithms is the training of models and often also predictions for unseen data. If a call to the training or prediction functions of a machine learning library instead yields unexpected exceptions, this is a clear indicator of a bug in the most crucial functions. 

We developed a set of smoke tests that we believe all machine learning algorithms must be able to pass. The tests are developed based on an equivalence class analysis of the input domain guided by the notion to identify problematic regions of the input domain that are likely to reveal bugs. Moreover, we consider that machine learning algorithms often have many hyperparameters. Hyperparameters can, e.g., configure the complexity of developed machine learning models (e.g., depth of a decision tree, number of neurons), but also configure the optimization algorithms (e.g., Newton's gradient descent or stochastic gradient descent) and, thus, imply that different parts of a software are executed. \cite{Chandrasekaran2017} found that consideration of hyperparameters is important for the coverage of tests. We show that exhaustive testing of hyperparameters with a grid search is not possible in general, due to the exponential nature of the problem. We propose a simple approach where the number of tests only grows linearly with the number of hyperparameters. 

We apply our approach to three state-of-the-art machine learning libraries to evaluate the effectiveness of our tests, i.e., if they are capable of detecting real-world bugs that were previously not detected. If we are able to find bugs in mature real-world software, it stands to reason that our smoke test suite is even more beneficial for any newly developed software. We discovered and reported eleven previously unknown bugs, two of which were critical bugs that can potentially crash not only the application running the machine learning tool, but also other applications in the same environment, because they trigger uncontrolled memory consumption. We found most of these bugs by testing with extremely large values close to machine precision. Moreover, some of the bugs also depend on the hyperparameters, which indicates that combinatorial testing is helpful. 

The main contributions of this article are the following.
\begin{itemize}
\item The definition of 22 smoke tests for classification algorithms and 15 smoke tests for clustering algorithms. 
\item A naive approach with linear complexity for combinatorial smoke testing that considers different values for all hyperparameters.
\item A case study in which we apply our methods to Weka~\citep{Frank2016}, scikit-learn~\citep{Pedregosa2011}, Spark MLlib to test 53 classification and 19 clustering algorithms that found eleven bugs that we reported to the developers.
\item A proof-of-concept that the approach also works with the deep learning framework TensorFlow~\citep{tensorflow2015-whitepaper}.
\item The recommendations of two best practices for developers about relatively simple and yet effective tests that can improve the quality assurance of machine learning libraries.
\end{itemize}

The remainder of this article is structured as follows. In Section~\ref{sec:related-work}, we discuss the related work on testing machine learning algorithms. Afterwards, we present our approaches for smoke testing in Section~\ref{sec:smoketests} and combinatorial smoke testing in Section~\ref{sec:combinatoric}. In Section~\ref{sec:experiments}, we apply our approach to real-world machine learning software and present the results, which we discuss in Section~\ref{sec:discussion}. We provide concrete recommendation for developers in Section~\ref{sec:recommendation}, before we conclude in Section~\ref{sec:conclusion}.

\section{Related Work}
\label{sec:related-work}

\cite{Zhang2020} and \cite{Braiek2020} recently published comprehensive literature reviews on software testing for machine learning. Both surveys did not identify best practices for smoke testing of machine learning or similar generic best practices for the testing of machine learning. Since our work can be considered as a combination of synthetic test input data generation and combinatorial testing, we discuss the related literature on such approaches in the following. Additionally, we consider approaches like fuzzing~\citep[e.g.][]{Xie2019} or code instrumentation to detect computational problems~\citep[e.g.][]{Bao2013} out of scope, as these are not basic techniques that can be setup with low effort for any machine learning algorithm. Other literature on testing of machine learning that, e.g., deals with the definition of test oracles with differential testing~\citep[e.g.][]{Pham2019} or metamorphic testing~\citep[e.g.][]{Murphy2008, Nejadgholi2019} are also out of scope. For a general overview on software testing for machine learning, we refer the reader to the aforementioned surveys. 

\subsection{Input Data for Machine Learning Testing}

According to the review by \cite{Zhang2020}, there are three publications that discuss the use of synthetic data for machine learning testing.\footnote{\cite{Zhang2020} also list \cite{Zhang2019} as a paper for synthetic test input generation. However, that paper does not address test input generation with the goal to test the correctness of the machine learning algorithm, but rather to test if the machine learning algorithm overfits the results, which, to our mind, is a different use case and related to model accuracy.} 

Similar to our work, \cite{Murphy2007a} defined data that may be problematic for machine learning with the goal to test corner cases, e.g., repeating values, missing values, or categorical data. The notable difference to our work is the scope: they only evaluated three ranking algorithms based on three criteria. Our proposed tests also consider repeating values (e.g., all zeros), as well as categorical data, but also many other corner cases, such as data close to machine precision. We do not consider missing values, because if and how such data should be handled is highly algorithm and implementation specific and, therefore, from our perspective not suitable for generic best practices regarding software tests. 

\cite{Breck2019} propose an approach to generate training data that tries to infer the schema of the data that is fed into learning algorithms. The automatically inferred schema is reported to the engineers who can validate that the inferred schema matches their expectations about the data. They also use this schema to generate new input data for testing, through randomized data generation such that the data matches the schema. This data is then fed into the learning algorithms to test if the schema description is sufficient or if the data schema may need to be made stricter to prevent faults at prediction time and instead enforce these faults at data validation time. Similar to our work, \cite{Breck2019} try to ensure that machine learning implementations provide the basic functions, i.e., do not fail given data. However, their approach is targeted specifically at applications that are already deployed and for which many different samples of training/test data are available, which can be used for schema inference. Thus, this approach is not applicable for general purpose machine learning libraries, which we target. 

\cite{Nakajima2016} describe that they generate linearly separable data as part of an approach for metamorphic testing. However, their work does not show how exactly the data is generated. The data the authors describe seems to mostly fit our most basic smoke test with uniform informative data. 

A different perspective on the test input generation was provided by \cite{Wang2020}. The authors consider trained neural networks as test input and describe how established and often used neural networks such as AlexNet \citep{Krizhevsky2017} can be modified to create new inputs for testing the prediction functions of libraries for the definition of neural networks. This perspective on test data is orthogonal to our work. We consider the input data used for training and predictions. With deep learning, the architecture of the network is technically a hyperparameter, but in practice the type of the model that should be learned (e.g., convolutional or recurrent). This leads to an infinite amount of possible ``algorithms'' that could be tested. Restricting this to architectures similar to those that are well known and used in practice, focuses the testing to consider ``algorithms'' that are more likely to be used in the future.

We could not identify additional related work in the review by \cite{Braiek2020}. \cite{Zhang2020} also discuss symbolic execution, search-based testing, and domain-specific input generation methods. However, such methods are not directly related to our work as they either require to generate test inputs dynamically based on prior test results and not statically prior to the test execution, or are target for specific algorithms and use cases. In comparison, our work aims at the description of smoke tests through input data that can be statically generated once and then included in a test suite without special effort in the tooling. 

\subsection{Combinatorial Machine Learning Testing}

To the best of our knowledge, there is only one publication so far on combinatorial testing of machine learning configurations through hyperparameters by \cite{Chandrasekaran2017}.\footnote{\cite{ma2019deepct} suggest a combinatorial approach to explore the input space in an adversarial manner to test if results are robustly correct. While this is combinatorial work, the authors do not consider hyperparameters, but rather features. Hence, this is rather a publication on input data generation for model robustness testing.} They state that using the default inputs does not cover all aspects of a classifier and that combinatorial testing is required to cover both valid and invalid hyperparameters of algorithms. They evaluate how grid search over hyperparameters affects the test coverage using input data from the UCI machine learning archive~\citep{Dheeru2017}. They observe that combinatorial testing increases the test coverage and that small data sets are sufficient to achieve a high test coverage. The two major limitations of the work by \cite{Chandrasekaran2017} are that they only analyze test coverage and mutation scores, instead of the capability to find real-world bugs and that they did not solve the problem of the exponential growth of tests with a grid search. Within our work, we demonstrate that combinatorial testing is important for the bug detection capabilities of a test suite and propose an approach with linear instead of exponential growth to bound the number of tests. 

\section{Smoke Tests}
\label{sec:smoketests}

Smoke tests assert ``that the most crucial functions of a program work, but not bothering with finer details''.\footnote{\url{https://glossary.istqb.org/search/smoke\%20test}} When we apply this definition to machine learning algorithms, the meaning depends on the type of algorithm. For example, classification algorithms have two crucial functions that must always work, given valid input: the training of the classification model and the prediction for instances given the trained model. The same holds true for regression algorithms. Clustering algorithms only have one crucial function, i.e., the determination of clusters in the training data. 

Within this article, we consider classification algorithms and clustering algorithms. Our general approach for smoke testing is simple: we feed valid data for training, and in case of classification also prediction, into the machine learning implementations and observe if the application crashes. The most basic criterion for smoke testing is that implementations must be able to handle common data distributions correctly. However, for effective smoke testing, data should also be chosen in such a way, that it is likely to reveal bugs and crash software. 

We identified eleven problems with data that algorithms should be able to handle correctly. We derived 21 smoke tests that address both unproblematic data, as well as the eleven problems with the data we identified. These smoke tests ensure that algorithms work on a basic level. Tables~\ref{tbl:smoketests-1} and~\ref{tbl:smoketests-2} list the motivation for a smoke test, the identifier, the way the features and the classification for the smoke tests are generated and our rational for these features/classes. Each smoke test was designed to test for a single problem within the data if possible to prevent error hiding. In the following, we discuss the equivalence classes and label generation in detail.

\begin{table}
\footnotesize
\begin{sideways}
\begin{tabular}{p{5.2cm}p{2cm}p{3.2cm}p{2.8cm}p{4.3cm}}
Motivation & Identifier & Features & Classification & Rational \\
\hline\hline
\multirow{2}{5.2cm}{Data with features in common data ranges and that is informative for the classification should not cause any problems.}
& UNIFORM & Uniform in [0, 1] & Rectangle & Normalized data with a common distribution. \\
& CATEGORICAL & Data with ten uniformly distributed categories & Rectangle & Informative categorical data with a reasonable number of categories. \\
\hline
\multirow{3}{5.2cm}{Extremely small values close to machine precision may lead to numeric underflows, e.g., if multiplied or overflows, e.g., if coefficients are calculated.}
& MINFLOAT & Uniform in [0, $10^{-6}$] & Rectangle & Machine precision for 32 bit floating point numbers.\\
& VERYSMALL & Uniform in [0, $10^{-10}$] & Rectangle & Between 32 bit and 64 bit floating point precision.\\
& MINDOUBLE & Uniform in [0, $10^{-15}$] & Rectangle & Machine precision for 64 bit floating point numbers.\\
\hline
\multirow{3}{5.2cm}{Extremely large values close to machine precision may lead to numeric underflows, e.g., if coefficients are calculated or overflows, e.g., if multiplied.}
& MAXFLOAT & Uniform in [0, $3.4\cdot10^{38}$] & Rectangle & Machine precision for 32 bit floating point numbers \\
& VERYLARGE & Uniform in [0, $10^{100}$] & Rectangle & Between 32 bit and 64 bit floating point precision.\\
& MAXDOUBLE & Uniform in [0, $1.7\cdot10^{308}$] & Rectangle & Machine precision for 64 bit floating point numbers \\
\hline
\multirow{2}{5.2cm}{Extreme probability distributions can cause problems, e.g., for optimization algorithms.}
& SPLIT & With 50\% probability uniform in [0, $10^{-5}$] and with 50\% probability uniform in [$10^{10}$, $10^{11}$] & Random & Algorithms may have problems with an unexplained large split between instance values.\\
& LEFTSKEW & Negative of a gamma distribution with $\kappa=0.1$, $\theta=4.0$ & Rectangle & Algorithms may have problems with a strong left skew. \\
& RIGHTSKEW & Gamma distribution with $\kappa=0.1$, $\theta=4.0$ & Rectangle & Algorithms may have problems with a strong right skew. \\
\hline
\multirow{2}{5.2cm}{Extreme class level imbalance, i.e., only very few data from one class.}
& ONECLASS & Uniform in [0, 1] & All data from the same class & Most extreme class level imbalance possible \\
& BIAS & Uniform in [0, 1] & All data from the same class, except one instance which is in the other class & Very strong class level imbalance.\\
\hline  
\end{tabular}
\end{sideways}
\caption{Description of possible problems with data for classification algorithms and derived smoke tests. The test data for the smoke tests is the same as the training data, unless specifically stated otherwise. (1/2)}
\label{tbl:smoketests-1}
\end{table}

\begin{table}
\footnotesize
\begin{sideways}
\begin{tabular}{p{5.2cm}p{2.5cm}p{3.2cm}p{2.8cm}p{4.5cm}}
Motivation & Identifier & Features & Classification & Rational \\
\hline\hline
Outliers, i.e., single instances with values far away from all other data. 
& OUTLIER & Uniform in [0, $10^{-5}$], except one instance with $10^{10}$ for all features & Rectangle & Algorithms may have problems with a single extreme outlier. \\
\hline 
Equal instances, i.e., data where all instances have exactly the same feature values.
& ZEROS & All values zero & Rectangle & In addition to all values equal, zeros may cause problems with divisions.\\
\hline 
\multirow{5}{5.2cm}{Random class assignments that do not allow separation of the classes, i.e., no features in the data carries any information about the classes. This may cause problems for feature selection or optimization algorithms used by classifiers.}
& RANDNUM & Uniform in [0, 1] & Random & Common feature values, but randomly assigned classes. \\
& RANDCAT & Categorical data with two classes & Random & Common number of categories, but randomly assigned classes. \\
& & & \\
& & & \\
& & & \\
\hline
Input for predictions that is not within the range of the observed training data.
& DISJNUM & Training data uniform in [0, 1], test data uniform in [100, 101] & Rectangle & Numeric values in training and test are disjunctive.\\
& DISJCAT & Training data with ten categories, test data with ten different categories & Rectangle & Categories in training and test are disjunctive.\\
\hline
Categorical data with many distinct categories.
& MANYCATS & 10000 categories for each feature & Rectangle & Large numbers of categories may inflate the dimension of the data. \\
\hline  
Categorical data with only single observations for at least one category.
& STARVEDMANY & Each instances gets a unique category value for each feature. & Random & Smallest number of samples possible for each category. \\
\hline
Categorical data where no data is available for some of the defined categories.
& STARVEDBINARY & Data with two defined categories in the meta data, such that the first feature only has data of the first category, the second feature only has data of the second category. & Rectangle & Smallest example with empty categories that also considers different category orders for the empty categories.\\
\hline 
\end{tabular}
\end{sideways}
\caption{Description of possible problems with data for classification algorithms and derived smoke tests. The test data for the smoke tests is the same as the training data, unless specifically stated otherwise. (2/2) }
\label{tbl:smoketests-2}
\end{table}

\subsection{Equivalence Classes}

Technically speaking, all smoke tests are defined over the same two equivalence classes, i.e., valid numeric and categorical data. Text books then propose to conduct a boundary-value analysis. What this means for machine learning algorithms is not obvious, because it is unclear where exactly the boundaries of the valid inputs are: we have multi-dimensional features as input, the differentiation between training and test data, and for classification algorithms also class labels. If we were to follow the theory, each of these input dimensions would be treated independently and we would analyse, e.g., the boundaries for each feature separately, while using unproblematic values for all other features and the class labels to prevent error hiding. However, such an approach would ignore that there are many interactions between these input parameters. Hence, we rather consider all these aspect as an ``input'' and our smoke tests define how each of these inputs should be defined to be a boundary of this complex input space.\footnote{We note that this boundary is unrelated to the decision boundary of models, because the quality of models is no concern of this work. Hence, work on testing how inputs close the decision boundary behave (e.g., \cite{Udeshi2018}) is out of scope.} Additionally, boundary value analysis is based on the idea that if, e.g., a loop works correctly for 0, 1, and the maximum number of possible iterations, it should also work for all numbers in between. Again, the considerations we have to make for differences force us to look at this differently, because not only absolute feature values matter, but also their probability distribution. For example, an algorithm that tries to fit a normal distribution might break at some point based on the shape of the input. This concept is not captured by the classic boundary value analysis. 

Based on these considerations, we rather shatter the valid equivalence class into multiple equivalence classes that cover different regions of the valid input. The different regions are ``boundaries'' in the sense that they are derived such that they represent ``extreme'' data that is likely to reveal bugs. For all equivalence classes the class labels are binary with roughly the same amount of data in each class, such that the features are informative (see Section~\ref{sec:labels}), unless we specifically state that this is different.

We identified one equivalence class that represents well conditioned data. 
\begin{itemize}
    \item Numeric data that is uniform or normally distributed and not close to machine precision, respectively categorical data where the number of categories is significantly larger than the amount of data and each category contains roughly the same amount of data. Such data is commonplace and should never lead to bugs. We derived the UNIFORM and CATEGORICAL tests from this equivalence class. 
\end{itemize}

The remaining equivalence classes were modeled such that they each represent different potentially problematic behaviors. Four equivalence classes deal with features that are likely to trigger problems due to precision problems with floating point arithmetic. 
\begin{itemize}
\item Floating point data close to zero in the area of with 32/64 bit precision. There is a risk of numeric underflows, which could, e.g., lead to divisions by zero. We derived the MINFLOAT, VERYSMALL, and MINDOUBLE tests from these equivalence classes. 
\item Floating point data close to the maximal possible values with 32/64 bit precision. There is a risk of numeric overflows, which could, e.g., lead to infinities or divisions by zero. We derived the MAXFLOAT, VERYLARGE, and MAXDOUBLE tests from these equivalence classes. 
\item Distributions with extreme outliers. Outliers could cause numeric problems, e.g., underflows of floating point operations due to the difference of the exponents in the floating point representation of these numbers. We derived the OUTLIER test this equivalence class. 
\end{itemize}

The next three equivalence classes consider cases of feature distributions which could be problematic for algorithms that infer distributional characteristics of the features.
\begin{itemize}
\item Distributions with a strong left skew or strong right skew. The long tail of the distribution can cause problems with the reliable estimation of distribution parameters, e.g., variances, which could become very large and cause problems. In extreme cases if may even be impossible to fit a distribution due to numeric problems. We derived the LEFTSKEW and RIGHTSKEW tests from this equivalence class.
\item Multimodal features with large gaps in the input space. The gap in the data may cause problems for algorithms that try to fit distributions to the data. These problems could lead to bad performance, which would not be detected by our smoke tests, but also to crashes in case no distribution at all can be fit, e.g., because the algorithm alternates between the different modes of the distribution without convergence. If this situation is not handled by defaulting to a baseline scenario or by limiting iterations this could lead to crashes. We derived the SPLIT equivalence class from this equivalence class.
\item Data with constant values. Such data could lead to problems due to unhandled numeric special cases, e.g., a variance of zero in an attempt to standardize the data. We derived the ZEROS test from this equivalence class.
\end{itemize}

We also identified three equivalence classes that consider special cases for categorical features. 
\begin{itemize}
\item The categories are chosen such that there is only one instance per category. This could, e.g., happen if an identifier is inadvertently used as categorical feature. This could result in problems when occurrence statistics are calculated. We derived the STARVEDMANY test from this equivalence class. 
\item All data is be in a single category, but the metadata indicates that there should be multiple categories. The single category could lead to problems, e.g., because data cannot be partitioned based on the features. we derived the STARVEDBINARY test from this equivalence class. 
\item There is a very large number of distinct categories. Many categories may lead to problems for algorithms that cannot handle categorical data naturally and instead, e.g., expand the categories into many binary features (one-hot encoding). This could lead to overflows and memory consumption problems, even if this is not warranted by the data size. We derived the MANYCATS test from this equivalence class. 
\end{itemize}

The class labels may also cause problems for classification algorithms. We identified two equivalence classes for problematic labels of classification data, that are independent of feature values. 
\begin{itemize}
\item All data from the same class. Algorithms that assume the presences of multiple classes without checking may crash with all sorts of numeric or indexing problems. For example, all calculations of distributional characteristics are impossible. Subsetting of data, stratified sampling, or other aspects may lead to index out of bounds problems or empty subsets due to the missing data. We derived the ONECLASS test from this equivalence class. 
\item Only a single training instance for a class. This could cause problems like divisions by zero during the calculation of the standard deviation of the values of a feature for that class. We derived the BIAS test from this equivalence class. 
\end{itemize}

We also identified one equivalence classes that consider the interactions between features and classes. 
\begin{itemize}
\item The features are all random and do not carry any information about the classes. An algorithm may fail to select any feature with relevance for the class which could result in empty models that must be handled. Moreover, the lack of information could lead to convergence problems of optimization algorithms because all solutions are equally bad, which could result in infinite loops. We derived the RANDNUM and RANDCAT tests from this equivalence class. 
\end{itemize}

Finally, we determined one equivalence class that is considering the relationship of training and test data. 
\begin{itemize}
\item The observed range of values of test data may differ from the training data. In case the prediction part of the algorithm makes assumptions about values, these may be violated by non-overlapping data. We derived the DISJNUM and DISJCAT tests from this equivalence class.
\end{itemize}

%Additionally, we defines one generic smoke test, that does not have an identifier. This smoke test is based on the idea, that if a data set once caused a classifier to crash, e.g., due to extreme distributions, skew, or similar, it is likely that other classifiers also have problems with the data. Thus, collecting such data is an easy way to define smoke tests. 

\subsection{Label Generation}
\label{sec:labels}

\begin{figure}
\centering
\includegraphics[width=\textwidth]{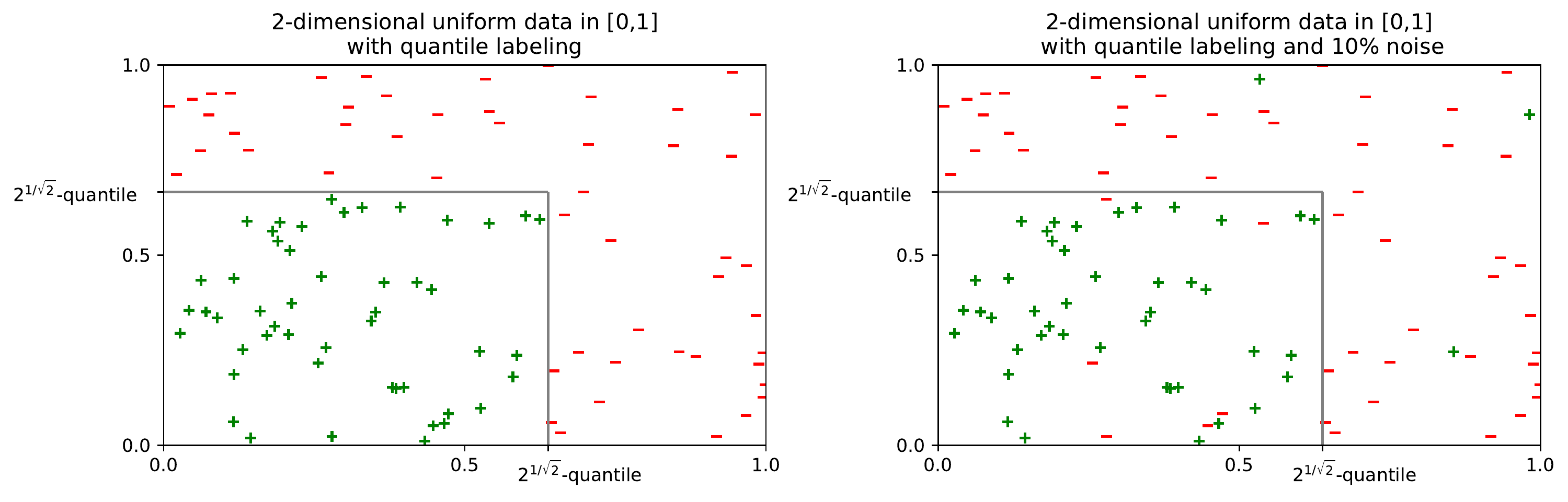}
\caption{Example for labeling in form of a rectangle using the quantiles.}
\label{fig:label_example}
\end{figure}

All classifications we generate are binary. There are two general strategies our smoke tests use: \emph{rectangle} and \emph{random}. With the rectangle labelling strategy we define an axis aligned rectangle such that roughly 50\% of the data belongs to each class. To achieve this for any probability distribution, we make use of the quantiles. Let $q$ be the $2^{-1/m}$-th quantile of a probability distribution $X$. Through the definition of quantiles, it follows that there is a $2^{-1/m}$ probability that a value drawn from $X$ is less than $q$, i.e., $P(X<q) = 2^{-1/m}$. If we randomly draw $m$ values from $X$, the probability that all values are less than $2^{-1/m}$ is $P(X<q)^m = 2^{-m/m} = 0.5$. After we label using quantiles, we apply 10\% white noise to the data, i.e., we flip labels with 10\% probability. Figure~\ref{fig:label_example} visualizes the rectangle labeling in two dimensions. Thus, data that is labelled using the rectangle strategy is informative, i.e., all features carry information about the class label. We note that the rectangle strategy can be replaced with any other approach that generates informative labels given data. However, the advantage of our approach is that it is independent of the distribution of the features. For the random labeling, we randomly assign class labels with 50\% probability, independent of the instance values. Thus, for data that is labelled using the random strategy, no feature carries any information about the class label. 

For the testing of clustering, we use the same data, but drop the labels and do not use the test data. Consequently, the tests RANDNUM, ONECLASS, BIAS, and DISJNUM are the same as UNIFORM for clustering, RANDCAT and DISJCAT is the same as CATEGORICAL. Hence, we only have 15 smoke tests for clustering.

\section{Combinatorial Smoke Testing}
\label{sec:combinatoric}

\cite{Chandrasekaran2017} found that it is important to test a broad set of hyperparameters of machine learning algorithms to achieve a good coverage of the tests. They further demonstrate that this coverage is correlated with a high mutation score, an estimation of the test suites capability to detect bugs. Therefore, our smoke tests would have limited efficiency, unless we apply them with different hyperparameters. Unfortunately, this problem is not trivial. We explain the concept of combinatorial testing of hyperparameters, its limitations, and how to overcome them, using the J48 decision tree classifier from Weka. Table~\ref{tbl:j48-params} lists the hyperparameters of the algorithm. Further, we note that hyperparameter optimization techniques that conduct an automated search of the hyperparameter space with the goal to find the best classifier, such as Bayesian optimization~\citep{Snoek2012} or gradient-based methods~\citep[e.g.,][]{Dougal2015} are not suitable for testing, as we do not intend to find the best solution, but rather to achieve a good coverage of hyperparameters.

A naive approach towards combinatorial testing would be exhaustive testing of all values. This would be $2^8 \cdot 3^2=2304$ hyperparameter combinations. Hence, the naive approach has exponential complexity. Thus, we would need to execute $21 \cdot 2304 = 48384$ test cases for the naive approach, just for one classifier. Such exhaustive combinatorial testing quickly requires huge amounts of computational resources. The high resource demand would counteract our goal to provide simple and generic tests that can be applied anytime to any algorithm that uses numeric or categorical input data. 

The second problem of the naive approach are invalid test cases, because not all parameter combinations are compatible with each other. For example, only one of the flags doNotMakeSplitPointActualValue and J may be used, as both specify how split points are determined for the creation of rules. Moreover, if the U flag is used to specify that the tree should not be pruned, the parameters C, S and R that configure the pruning may not be used. The parameter N is specific for REP and may only be used if the flag R is used. Vice versa, the parameter C is specific for pruning without REP and may not be used if R is used. 

Thus, a naive approach would not only lead to a large computational complexity, but also many invalid test cases due to incompatible hyperparameters. We propose two modifications to achieve linear complexity and avoid invalid test cases: only modify one parameter at a time and use sets that define valid combinations. 

We achieve linear complexity by taking pattern from test coverage criteria. For us, the naive combinatorial approach is similar to path coverage, i.e., executing all possible paths of executions. Similar to our problem, this problem grows exponentially and is, therefore, not a reasonable test coverage criterion for many applications. A weaker, but still very useful criterion is the condition coverage: all atomic conditions assume all possible values (i.e., true and false) at least once. We apply this idea to the hyperparameters: instead of looking at all possible combinations, we use default values for all but one parameter, which we modify. The different values for the parameters should be used such that they cover all equivalence classes. Ideally, multiple values from each equivalence class are used, such that inputs at the boundaries of the equivalence classes are used, similar to a boundary-value analysis. 

This approach leads to a number of test cases that is linear in the number of hyperparameters. Moreover, we can use this concept to partially solve the problem of invalid tests cases: if we use the default value disabled for doNotMakeSplitPointActualValue and J, they will never be both enabled.  However, the other problems with invalid tests cannot be solved that way: if multiple parameters are only allowed in specific combinations, we have to prescribe these conditions. We achieve this by not considering the hyperparameters of J48 as a single set, but as three sets: 
\begin{itemize}
    \item In the first set, the flag U is always enabled and the flag R is always disabled. This set represents the case of unpruned trees and tests combinations of the parameters M, A, doNotMakeSplitPointActualValue, J, and O. This leads to 6 hyperparameter combinations. 
    \item In the second set, the flags U and R are both always disabled. This set represents pruning without REP and tests combinations of the parameters M, A, doNotMakeSplitPointActualValue, J, O, S, and C. This leads to 11 hyperparameter combinations. 
    \item In the third set, the flag U is always disabled and the flag R is always enabled. This set represents pruning with REP and tests combinations of the parameters M, A, doNotMakeSplitPointActualValue, J, O, S, and N. This leads to 11 hyperparameter combinations.
\end{itemize}

Overall, we have $6+11+11 = 28$ hyperparameter combinations for this classifier leading to $21 \cdot 28 = 588$ test cases, i.e., only 1.2\% of the combinations of the naive approach, while still testing a reasonable set of hyperparameters. 

\begin{table}[]
\footnotesize
\centering
\begin{tabular}{p{2cm}p{5.7cm}lp{1.3cm}}
\textbf{Parameter} & \textbf{Meaning}  & \textbf{Default} & \textbf{Values} \\
\hline\hline
M & Integer that defines the minimum number of instances required for nodes in a tree. & 1 & 1, 10\\
A & Flag that defines if Laplace correction is used. & Disabled  & Enabled, Disabled \\
doNotMakeSplit-PointActualValue & Flag that forces split point of rules at values observed in the training data. & Disabled & Enabled, Disabled \\
J & Flag that defines if Minimum Description Length (MDL) correction is used for the split points of rules. & Disabled & Enabled, Disabled \\
O & Flag that defines if parts of the tree may be removed, if they do not reduce the training error. & Disabled & Enabled, Disabled \\
U & Flag that defines if the decision tree will be pruned at the end of the training to avoid overfitting. & - & - \\
S & Flag that defines if sub-trees may be raised as part of the pruning process. & Disabled & Enabled \\
C & Confidence factor for the pruning of the tree at the end of the training. & 0.25 & 0.05, 0.5, 0.95 \\
R & Flag that defines if Reduced Error Pruning (REP) is used. & - & - \\
N & Integer that defines the number of cross-validation folds used for REP. & 3 & 2, 3, 4\\
\hline
\end{tabular}
\caption{Hyperparameters of the J48 decision tree classifier of Weka. We note that there are additional hyperparameters, e.g., for nominal data, the numeric precision of results, debugging, or the definition of the seed for the random number generated. "-" marks parameters that have fixed values based on the parameter set. }
\label{tbl:j48-params}
\end{table}

\section{Case Study}
\label{sec:experiments}

We conducted a case study using three machine learning libraries as subjects to analyse the effectiveness of our tests. Within this section, we describe the subjects we selected, our implementation of the tests, as well as the results of the test execution. A replication package for the case is provided online.\footnote{https://github.com/sherbold/replication-kit-2020-smoke-testing - A DOI citable Zenodo archive will be created in case of acceptance.}

\subsection{Subjects}

We used purposive sampling for our case study~\citep{Patton2014} based on three criteria. First, we restricted the subjects to mature machine learning libraries that already have many users and should, therefore, already have a high quality, especially with respect to the basic functions to which we apply the smoke testing. Hence, we selected subjects for which we believe that finding bugs is especially hard. Our hypothesis is that if our tests are effective for these libraries, this means that the tests are also effective for other libraries and machine learning algorithms. Second, we require that all subjects are implemented in different programming languages and by, to the best of our knowledge, independent project teams. This should ensure that our results generalize beyond the social-technical aspects of a specific project. Third, the subjects should cover a broad selection of algorithms and not be specialized for a single purpose. The libraries must at least cover supervised learning with classification algorithms and unsupervised learning with clustering. This criterion ensures that we can evaluate a broad set of algorithms within our case study. We selected three such libraries. For all libraries, we used the latest published release for testing.\footnote{The tests were run in July 2020.} All bugs were verified against the latest commit of the main branch of the repository at the time we reported them. 

\begin{itemize}
\item Weka~\citep{Frank2016}: Weka is a popular machine learning library for Java which had the first official release in 1999. Thus, the library is very mature and
has been used by researchers and practitioners for decades and is, e.g., part of the Pentaho business intelligence software.\footnote{https://www.hitachivantara.com/en-us/products/data-management-analytics/pentaho-platform.html} All tests were executed against version 3.9.2 of Weka. Table~\ref{tbl:algs-weka} lists the 23 classification and six clustering algorithms that we tested. These are all algorithms that distributed together with Weka releases, i.e., that are not developed in separate packages, with the exception of meta learners.\footnote{We excluded meta learners from our study, because they internally use another classifier, i.e., they often propagate the learning to other classifiers. Thus, meta learners require special consideration, especially with respect to their hyperparameters, which are out of scope of out work.} 
\item scikit-learn~\citep{Pedregosa2011}: scikit-learn is a popular machine learning library for Python and one of the reasons for the growing popularity of python as a language for data analysis. All tests were executed against version 0.23.1 of scikit-learn with numpy version 1.15.0\footnote{numpy is used for mathematical calculations by scikit-learn.}. Table~\ref{tbl:algs-sklearn} lists the 23 classification and ten clustering algorithms that we tested. These are all non-experimental algorithms available in scikit-learn, with the exception of meta learners. 
\item Spark MLlib: Spark MLlib is the machine learning component of the rapidly growing big data framework Apache Spark that is developed with Scala. All tests were executed against the version 3.0.0 of Spark MLlib. Table~\ref{tbl:algs-spark} lists the seven classification and three clustering algorithms that we tested. These are all non-experimental algorithms, that are shipped together with the MLlib.
\item TensorFlow~\citep{tensorflow2015-whitepaper}: TensorFlow is one of the major frameworks for deep learning. All tests were executed against TensorFlow 2.3.0. Due to the number of possible ways that neural networks could be configured (layer types, activation functions, and their combinations) and trained (loss functions, stopping criteria, optimization algorithms), an exhaustive exploration of these frameworks is out of the scope of this article. Instead, we only test a multi-layer perceptron with the same options as the MLPClassifier of scikit-learn as proof-of-concept.
\end{itemize}

\begin{table}
\footnotesize
\begin{tabular}{lrrr}
\textbf{Identifier} & \textbf{\#Parameters} & \textbf{\#Combinations} & \textbf{\#Exhaustive} \\
\hline\hline
J48 (C4.5) & 10 & 28 & 2,304 \\
DecisionStump & 0 & 1 & 1 \\
HoeffdingTree & 6 & 13 & 729 \\
LMT & 9 & 12 & 1,728\\
RandomForest & 9 & 13 & 1,728 \\
RandomTree & 5 & 9 & 108 \\
REPTree & 7 & 12 & 972 \\
DecisionTable & 4 & 9 & 60 \\
Ripper & 5 & 8 & 108 \\
OneR & 1 & 3 & 3 \\
PART & 8 & 22 & 865 \\
ZeroR & 0 & 1 & 1 \\
IBk (KNN) & 6 & 11 & 240 \\
KStar & 3 & 6 & 24 \\
BayesNet & 3 & 7 & 20 \\
NaiveBayes & 2 & 3 & 4 \\
NaiveBayesMultinomial & 0 & 1 & 1 \\
Logistic & 2 & 3 & 6 \\
MultilayerPerceptron & 11 & 24 & 96,768 \\
SGD & 6 & 12 & 432 \\
SimpleLogistic & 7 & 11 & 648 \\
SMO (SVM) & 7 & 12 & 432 \\
VotedPerceptron & 3 & 6 & 27 \\ % sum for weka classifiers: 224 vs. 107209 (10441 ohne MLP)
\hline
Canopy & 7 & 14 & 1,458 \\
Cobweb & 3 & 7 & 24 \\
EM & 10 & 18 & 17,496 \\
FarthestFirst & 1 & 3 & 3 \\
HierarchicalClusterer & 5 & 15 & 480 \\
SimpleKMeans & 15 & 21 & 2,820,096 \\
\hline
\end{tabular}
\caption{Algorithms from Weka that we tested. \#Parameters are the hyperparameters of the algorithm that we considered. \#Combinations is the number of combinations of parameters we derived for our testing. \#Exhaustive is the number of combinations we would have, without our bounding strategy. Above the separator are classification algorithms, below are clustering algorithms. }
\label{tbl:algs-weka}
\end{table}

\begin{table}
\footnotesize
\begin{tabular}{lrrr}
\textbf{Identifier} & \textbf{\#Parameters} & \textbf{\#Combinations} & \textbf{\#Exhaustive} \\
\hline\hline
DecisionTreeClassifier & 11 & 24 & 163,296 \\
ExtraTreeClassifier &  11 & 24 & 163,296 \\
RandomForestClassifier & 12 & 27 & 734,832 \\
DummyClassifier & 2 & 6 & 10 \\
GaussianProcessClassifier & 4 & 5 & 18 \\
PassiveAggressiveClassifier & 10 & 14 & 2,592 \\
RidgeClassifier & 8 & 16 & 1,344 \\
SGDClassifier & 15 & 29 & 2,099,520 \\
BernoulliNB & 3 & 6 & 18 \\
CategoricalNB & 2 & 4 & 6 \\
ComplementNB & 3 & 5 & 12 \\
MultinomialNB & 2 & 4 & 6 \\
KNeighborsClassifier & 6 & 12 & 648 \\
RadiusNeighborsClassifier & 7 & 16 & 2,592 \\
MLPClassifier & 18 & 33 & 45,349,632 \\
LinearSVC & 9 & 12 & 864 \\
NuSVC & 7 & 28 & 576 \\
SVC & 8 & 29 & 864 \\
LinearDiscriminantAnalysis & 5 & 11 & 60 \\
QuadraticDistriminantAnalysis & 3 & 4 & 6 \\ % sum for sklearn classifiers: 309
GradientBoostingClassifier & 15 & 31 & 14,880,348 \\ 
Perceptron & 13 & 20 & 93,312 \\
NearestCentroid & 1 & 2 & 2 \\
\hline
AffinityPropagation & 7 & 11 & 432 \\
AgglomerativeClustering & 5 & 19 & 648 \\
Birch & 4 & 8 & 54 \\
DBSCAN & 7 & 21 & 11,664 \\
KMeans & 10 & 20 & 29,160 \\
MiniBatchKMeans & 10 & 19 & 31,104 \\
MeanShift & 4 & 7 & 31104 \\
OPTICS & 12 & 30 & 1,119,744 \\
SpectralClustering & 12 & 40 & 209,952 \\
GaussianMixture & 9 & 17 & 7,776 \\
\hline
\end{tabular}
\caption{Algorithms from scikit-learn that we tested. \#Parameters are the hyperparameters of the algorithm that we considered. \#Combinations is the number of combinations of parameters we derived for our testing. \#Exhaustive is the number of combinations we would have, without our bounding strategy. Above the separator are classification algorithms, below are clustering algorithms. }
\label{tbl:algs-sklearn}
\end{table}

\begin{table}
\footnotesize
\begin{tabular}{lrrr}
\textbf{Identifier} & \textbf{\#Parameters} & \textbf{\#Combinations} & \textbf{\#Exhaustive} \\
\hline\hline
DecisionTreeClassifier & 5 & 10 & 162 \\
RandomForestClassifier & 8 & 22 & 18,144 \\
GBTClassifier & 8 & 18 & 2,268 \\
LogisticRegression & 8 & 14 & 2,916 \\
MultilayerPerceptronClassifier & 4 & 8 & 54 \\
LinearSVC & 6 & 10 & 324 \\
NaiveBayes & 2 & 10 & 12 \\
\hline
KMeans & 6 & 11 & 324 \\
GaussianMixture & 3 & 7 & 27 \\
BisectingKMeans & 3 & 7 & 27 \\
\hline
\end{tabular}
\caption{Algorithms from Apache Spark that we tested. \#Parameters are the hyperparameters of the algorithm that we considered. \#Combinations is the number of combinations of parameters we derived for our testing.  \#Exhaustive is the number of combinations we would have, without our bounding strategy. Above the separator are classification algorithms, below are clustering algorithms. }
\label{tbl:algs-spark}
\end{table}

\subsection{Implementation of Tests}

We implemented a prototype of the combinatorial smoke testing in the prototype \atoml{}\footnote{https://github.com/sherbold/atoml}. The goal of our implementation was to demonstrate that the combinatorial smoke testing can be automated to a large degree, even if sets of valid hyperparameters must be specified by developers. \atoml{} generates training data with the specified numbers of features and samples using the random number generators of Apache Commons Math.\footnote{\url{https://commons.apache.org/proper/commons-math/}} The class labels are generated as a categorical variable. Table~\ref{tbl:data_example} shows examples of data generated for the UNIFORM and RIGHTSKEW tests. The foundation for tests for specific frameworks is a template engine that \atoml{} provides. A template is specific to a framework and algorithm type, e.g., Weka classification algorithms, or scikit-learn clustering algorithms. Currently, \atoml{} supports all classification and clustering algorithms from the machine learning frameworks Weka, scikit-learn, and Apache Spark. 

\begin{table}[]
\centering
\begin{minipage}{0.45\textwidth}
\centering
\begin{tabular}{rrr}
\textbf{feature\_1} & \textbf{feature\_2} & \textbf{class} \\
\hline
0.072777 & 0.334995 & class\_0 \\
0.857577 & 0.977991 & class\_1 \\
0.310364 & 0.230206 & class\_1 \\
0.75821 & 0.600593 & class\_1 \\
0.883202 & 0.066408 & class\_0 \\
\end{tabular}

\vspace{4pt}
a) UNIFORM
\end{minipage}
\begin{minipage}{0.45\textwidth}
\centering
\begin{tabular}{rrr}
\textbf{feature\_1} & \textbf{feature\_2} & \textbf{class} \\
\hline
0.2 & 3.96107 & class\_0 \\
0.000048 & 0.624823 & class\_0 \\
0.000002 & 5.509755 & class\_1 \\
13.591907 & 8.23233 & class\_1 \\
0 & 0.000001 & class\_0 \\
\end{tabular}

\vspace{4pt}
b) RIGHTSKEW
\end{minipage}
\caption{Examples for test data with two features and five instances generated by \atoml{}.}
\label{tbl:data_example}
\end{table}

Algorithm specific parts are defined by a YAML dialect for the test specification. These test specification must be provided by the users of \atoml{} and are the only aspect that is not automated. Listing~\ref{lst:j48} shows how the parameter set for Weka's J48 classifier would look like with atoml. The description starts with the general data about the algorithm that should be tested. The name is used as prefix for the generated test cases. The type and framework define which template is used. The package and class specify where the algorithm under test is implemented. The features define the valid feature types for the algorithm under test, e.g., categorical features or numeric feature with double precision. This can be used to restrict the smoke tests, e.g., by not allowing categorical features. 

Finally, the parameters section defines the hyperparameters for the algorithm. All parameters have a default value. This value is used, when the possible hyperparameters for other parameters are tested. In case only a default value is specified, this means that all tests in the parameter set use the same value for that parameter. To determine which values for the hyperparameters should be considered, \atoml{} supports four types: flags, integers, floats, and value lists. Flags can either be enabled and disabled. \atoml{} uses flags instead of booleans, because of the command-line style API of Weka. For libaries that use booleans, the enabled flag is translated to true and disabled to false. The values tested for integers and floats are defined using ranges, that are specified by their min, max, and step size. All values in that range will be used for the smoke tests. Value lists can be used to specify any kind of parameter type, including strings. 

The tests that \atoml{} generates are in standard unit testing frameworks of the respective programming language, e.g., JUnit for Java, and unittest for Python. Thus, test suites generated by \atoml{} can be integrated in existing test suites of the libraries without any effort. Thus, while our prototype is not production ready (e.g, due to a lack of integration in build systems for the test generation), our implementation demonstrates that the concept can be used in practice without any major restrictions, e.g., on the machine learning framework, programming language, or type of algorithm. 

\begin{lstlisting}[float, language=yaml, caption=Definition of the J48 without pruning for \atoml{}., label=lst:j48]
name: WEKA_C45_UNPRUNED
type: classification
framework: weka
package: weka.classifiers.trees
class: J48
features: [double,categorical]
parameters:
  U: # only default means that this will always be enabled
    default: enabled
  M:
    type: integer # can sample over ranges
    min: 1
    max: 10
    stepsize: 9
    default: 2
  O:
    type: flag
    default: disabled
  A:
    type: flag
    default: disabled
  doNotMakeSplitPointActualValue:
    type: flag
    default: disabled
  J:
    type: flag
    default: disabled
\end{lstlisting}

\subsection{Results}

We ran all smoke tests defined in tables~\ref{tbl:smoketests-1} and~\ref{tbl:smoketests-2} against all algorithms under test and found problems in all three subjects. Overall, we found and reported the following eleven bugs. We note that one of the bugs affected seven algorithms. 

\begin{enumerate}
\item The Weka classifiers J48, LMT, HoeffdingTree, RandomForest, RandomTree, PART, and REPTree may all cause a stack overflow or running out of heap space.\footnote{https://jira.pentaho.com/browse/DATAMINING-779} We found this problem through the MAXDOUBLE test. The underlying problem seems to be an infinite recursion in the training, due to an unchecked bad numerical comparison that is used by all of these algorithms. The type of the exception depends on whether the JVM first runs out of stack or heap space, i.e., on the JVM configuration. We found this bug with all hyperparameter combinations we tested. 
\item The Weka classifier BayesNet may crash due to negative array indices caused by an integer overflow.\footnote{https://jira.pentaho.com/browse/DATAMINING-780} We found this problem with the STARVEDMANY test. This bug seems to be caused by the combination of the local search method and the identification of the structure of the Bayesian network. This bug was only triggered by five of the seven hyperparameter combinations we used and depending on how the parameters configured the local search on the configuration of the local search. 
\item The Weka classifier SMO may crash due to NaNs.\footnote{https://jira.pentaho.com/browse/DATAMINING-781} We found this problem through the MAXDOUBLE test. The problem was caused by trying to standardize the data, which failed due to the unchecked bad numerical conditioning. This bug occurs only if the standardization is selected, if normalization or no pre-preprocessing are selected through the hyperparameters, the problematic standardization code is not executed. 
\item The Weka clusterer EM may crash due to an unchecked estimation of the number of clusters and starting distributions through the internally used SimpleKMeans clusterer.\footnote{https://jira.pentaho.com/browse/DATAMINING-782} We found this problem through the MAXDOUBLE test. This problem seems to occur in corner cases with extreme data and only happens this way, because EM clustering supports -1 clusters (automatically pick number of clusters), while SimpleKMeans requires that the number of clusters is a positive integer. The extreme data causes EM not to pick a default number of clusters and instead to pass the -1 to SimpleKMeans, causing the crash. We found this bug with all hyperparameter combinations we tested. 
\item The Weka HierarchicalClusterer may crash due to an array index out of bounds.\footnote{https://jira.pentaho.com/browse/DATAMINING-783} We found this problem with the MAXDOUBLE test. The bug is caused by a lack of robustness of the distance calculations within the FilteredDistance approach that can be configured. This leads to instances not being assigned to any cluster, which is not allowed with hierarchical clusters. This bug can only be found, if the FilteredDistance is configured through the hyperparameters and was triggered by only one of the 15 hyperparameter combinations we used. 
\item  The scikit-learn AgglomerativeClustering could cause a MemoryError.\footnote{https://github.com/scikit-learn/scikit-learn/issues/17960} We found this problem through the MAXDOUBLE test. The bug was caused by a bad graph exploration strategy that led to revisiting highly connected nodes in the single linkage clustering many times, which consumed too much memory. The bug only happens if single linkage is selected through the hyperparameters and was triggered by only one of the 19 hyperparameter combinations. 
\item The scikit-learn classifiers MultinomialNB and ComplementNB can crash with an array index out of bounds if all data is from a single class.\footnote{https://github.com/scikit-learn/scikit-learn/issues/17926} We found this problem through the ONECLASS test. The bug is caused by a missing safeguard in the MultinomialNB probability calculation for the likelihoods in case a single class is used. The bug only occurs if the hyperparameters configure that the class prior is not fit and was triggered by only one of the four parameter combinations we used.
\item The scikit-learn classifier NearestCentroid can crash if all values features have the same values.\footnote{https://github.com/scikit-learn/scikit-learn/issues/18324} We found this problem through the ZEROS test. The bug seems to be caused by an uncontrolled shrinkage, that does not allow predictions with such data. The bug only occurs if the hyperparameter for shrinking features is, i.e., only one of the two hyperparameter combinations.
\item The scikit-learn clusterer Birch may crash with an attribute error because the value of an attribute is not calculated.\footnote{https://github.com/scikit-learn/scikit-learn/issues/17966} We found this problem trough the MAXDOUBLE test. The bug is caused by an overflow in an internal calculation where the values cannot be represented as floating point numbers anymore. We found this bug with all hyperparameter combinations we tested. 
\item The scikit-learn RidgeClassifier may crash with a numeric overflow.\footnote{https://github.com/scikit-learn/scikit-learn/issues/17925} We found this problem through the MAXDOUBLE test. Same as for the bug in the Birch clusterer, the cause is a numeric overflow. We found this bug with all hyperparameter combinations we tested.
\item The Apache Spark clusterer GaussianMixture may crash with an internal error.\footnote{https://issues.apache.org/jira/browse/SPARK-32569} We found this problem through the MAXDOUBLE test. The bug seems to be caused by an eigenvalue decomposition that does not converge. We found this bug with all hyperparameter combinations. 
\end{enumerate}

Additionally, our tests triggered several exceptions, in which our tests would have found bugs, but where the special cases were already actively handled by the developers through exceptions. While these are not bugs for the machine learning libraries we test, we note that none of these exceptions is part of the API documentation. Consequently, badly conditioned data during re-training of a production system based on these libraries could fail unexpectedly during operation. 

\begin{itemize}
\item The Weka clusterer SimpleKMeans crashes with an illegal argument for the ZEROS tests. The reason is the attempt to normalize feature values which fails because their sum is zero. This issue is only triggered if the hyperparameter for initialization is set to use $k$-Mean++. Because the exception indicates the exact problem, we give the benefit of doubt and do not classify this as bug. However, a fall back solution without normalization should also be possible with $k$-Means++ initialization.
\item The scikit-learn classifier MLPClassifier, PassiveAggressiveClassifier, QuadraticDiscriminantAnalysis, and SGDClassifier all throw exceptions for the for BIAS test that indicate that more than one instance is required for each class. While this happens for all hyperparameters of the QuadraticDiscriminantAnalysis, the other classifier have no problems with BIAS, unless early stopping is activated.
\item The scikit-learn classifiers GaussianProcessClassifier, LinearSVC, NuSVC all throw exceptions for the ONECLASS test that indicate that data for a single class is not allowed. 
\item The scikit-learn classifier DecisionTreeClassifier, ExtraTreeClassifier, and RandomForestClassifier directly specify that they only support values within the range of 32 bit floating number. We note that the algorithms have no problem with the VERYLARGE test, which are also outside of the 32 bit range. 
\item The scikit-learn classifier SGDClassifier reports problems due to numeric under-/overflows for the MAXDOUBLE smoke test and recommends rescaling the data. 
\item The scikit-learn classifier QuadraticDiscriminantAnalysis fails to converge for an internal singular value decomposition for the MAXDOUBLE test. Since non-convergence is always possible, we are not sure if this is a bug in the QuadraticDiscriminantAnalysis, a bug in the underlying implementation of the singular value decomposition, or if the matrix is really badly conditioned. Further investigation is still ongoing. 
\end{itemize}

The tests for TensorFlow did not reveal any errors. However, this is not surprising as we only used a very small part of the framework.

\begin{figure}
\centering
\includegraphics[width=\textwidth]{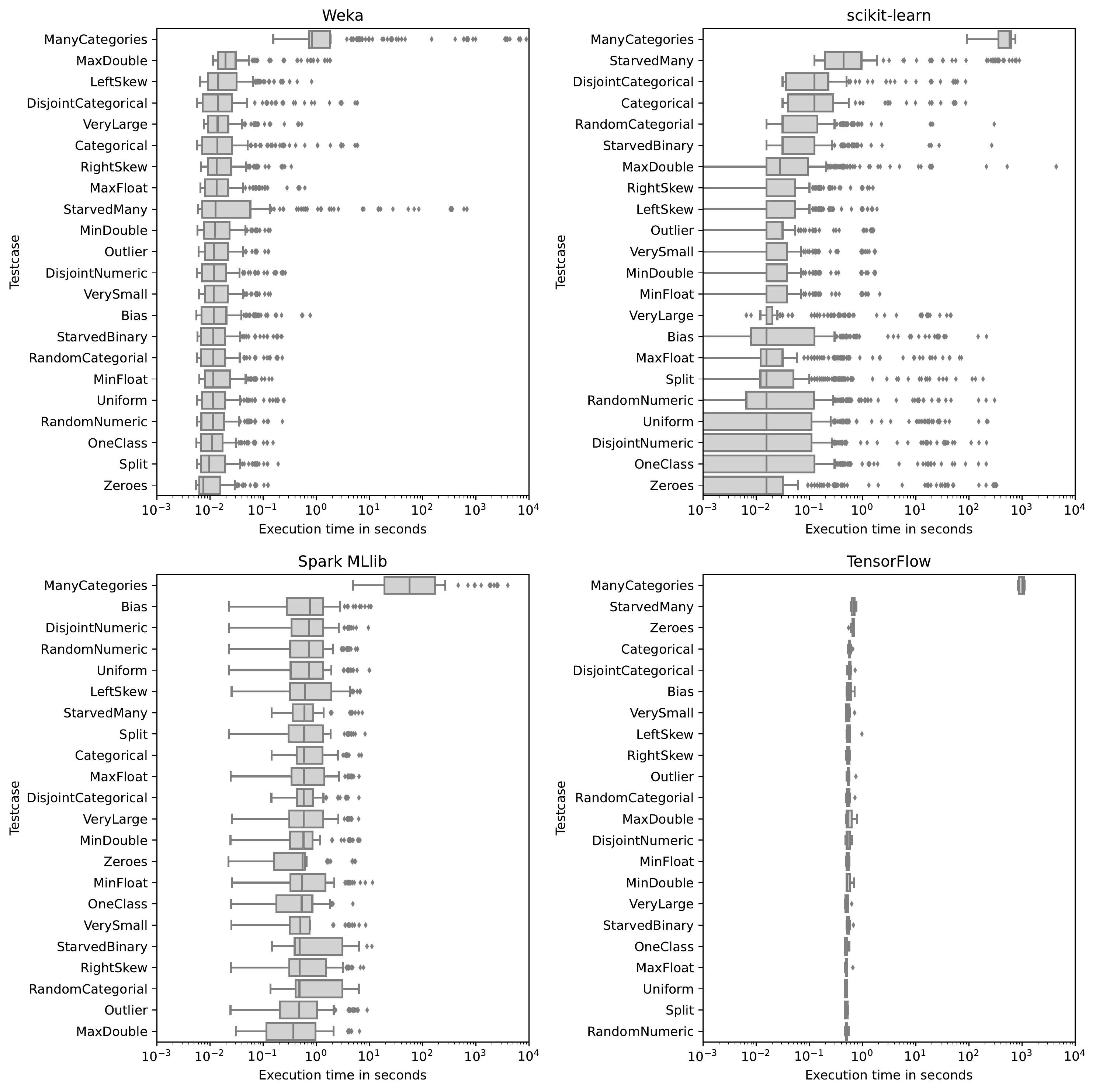}
\caption{Execution times of the tests we executed.}
\label{fig:execution-times}
\end{figure}

We measured the execution time of the tests to provide an indication of the computational resources required to conduct the testing.\footnote{We used a machine with Windows 10 Pro 21H1, Intel(R) Core(TM) i5-4570 CPU @ 3.20GHz, and 16 Gigabyte of memory for the measurements.} Figure~\ref{fig:execution-times} summarizes the test execution times. Overall, we required about 93 hours to run all tests once. About 92\% of them ran for less than one second. The exception to this short execution time are the MANYCATS tests, which required 79 hours of the overall execution time. However, this longer execution time in comparison to the other tests is expected, because this test is designed to induce stress through a large data set with many distinct categories. For the other tests, there is only a relatively low dispersion in the execution time, which is also expected as all tests use the same amount of features and instances as input for the algorithms. The remaining dispersion is due to the differences in the execution time of different algorithms. Consequently, there is almost no dispersion for TensorFlow, where we only tested a single algorithm.

\section{Discussion}
\label{sec:discussion}

We now discuss the results of our case study with respect to the effectiveness of our tests, the capability to find severe bugs, the need for combinatorial testing, and the generalizability of our results. We then derive an answer for our underlying research question, i.e., if we were successful in defining simple, generic, and effective tests for machine learning algorithms. 

\subsection{Effectiveness and Efficiency}

The effectiveness of our tests varies strongly. Only four of the 22 smoke tests found bugs (MAXDOUBLE, STARVEDMANY, ONECLASS, ZEROS). However, we also found exceptions that explicitly exclude additional problematic corner cases (MINDOUBLE, VERYSMALL, BIAS), which indicates that the correct handling of these corner cases should be part of test suites. Especially the MAXDOUBLE smoke test seems to be effective and can uncover bugs that only happen with extreme data. Interestingly, the smoke tests with data close to the lower bound of machine precision caused fewer problems. We believe this may be the case because numeric underflows often lead to zeros, which only cause problems when used in a division, while numeric overflows usually lead to NaNs or infinities, which lead to problems in all arithmetic operations. However, this could also be due to our subjects and not generalizable. The same holds true for the other smoke tests that did not uncover any bugs: while this is certainly an indication that these tests may not be effective at all, it may also be that we did not find any bugs because of our selection of subjects. 

Regarding the efficiency, we found that all tests, with the exception of MANYCATS run very fast and can be executed efficiently. Since MANYCATS did not find any bugs, we believe that this test could be dropped from our smoke test suite with a low risk. Developers can easily run the other smoke tests locally, whenever they modify an algorithm. Still, due to the large number of algorithms and many hyperparameters, an extensive test campaign that runs all tests for all algorithms may still require several hours for the execution, depending on the framework. This may be too expensive to be executed by a continuous integration system on every commit, but could instead be done nightly or weekly.

\begin{framed}
At least the smoke tests MAXDOUBLE should always be used and STARVEDMANY, ONECLASS, and ZEROS are also effective for corner cases. Further evidence indicates that MINDOUBLE, VERYSMALL, and BIAS may also be effective, but our evidence for this is only indirect, because the developers actively prevented the problems we expected. While single algorithms can be tested locally very fast, a full test campaign for all algorithms of a large framework should not be executed for every build, but rather with a fixed schedule, e.g., nightly.
\end{framed}

\subsection{Severity of the Bugs}

A potential problem of our approach is that if we identify almost trivial tests and avoid using a test oracle other than that the application crashes, we may only find trivial bugs. However, our results indicate that this is not the case. The bugs (1) and (6) are both major issues that could potentially even crash other applications running on the same machine, because both lead to the consumption of all available memory.

We already got substantial feedback from the developers regarding our bugs. While a common theme is that these are corner cases, that may not be relevant to many users, they usually acknowledged that the current behavior is problematic, regardless. Three of the bugs we reported for scikit-learn are already fixed. The discussion on how to best address the underlying issue is still ongoing for the fourth bug. Moreover, bug (9) was linked to a problem that was reported over three years ago\footnote{https://github.com/scikit-learn/scikit-learn/issues/6172} that the developers could never completely solve. The Weka developers acknowledged all bugs on the mailing list and are open for patches to address them.\footnote{We contacted them via Email after we realized that the Jira issue tracker is no longer in use since the company Pentaho was sold to Hitachi.} The only bug that the developers decided is not worth addressing is (11), which was marked by the Apache Spark developers as \texttt{Won't Fix}. 

A criticism of our statement that these bugs are indeed severe may be that such data is uncommon and may not occur in practice. While this may be correct, this ignores the potential threat of malign actors who intentionally poison data with uncommon values, e.g., with the goal to conduct denial of service attacks. Thus, if such bugs surface in end-user applications, the consequences can be severe. The best solution is to prevent bugs in libraries directly, which is why we conducted our research on this level. However, there is no reason that our tests can and should not be used on application level as well, to ensure that inputs are properly validated and handled. Anecdotally, we would like to add that this potential consequence was not raised by us, but rather by the host of developer podcast\footnote{https://fuzzyquality.com/episode/fuzzy-quality-episode-4-smoke-testing-machine-learning} that discussed an earlier version of this article with us.

\begin{framed}
The smoke tests are able to find non-trivial bugs that can be very severe, especially if exploited by malign actors.
\end{framed}

\subsection{Impact of Combinatorial Testing}

We could have missed six bugs if we would not have used a combinatorial approach that considered different values for all hyperparameters. Therefore, our results are a clear indicator that combinatorial testing over hyperparameters is important. This conclusion is further supported by the prior work from \cite{Chandrasekaran2017} that found that combinatorial testing is important for test coverage. We also find that our approach for pruning the number of combinatorial tests still lead to effective tests that found bugs, i.e., we could drastically reduce the number of hyperparameter combinations we need to consider, while still finding bugs. We note that the combinatorial testing is still naive and could be further improved, e.g., by considering interactions between parameters with pairwise testing~\citep{Kuhn2004} or by considering algorithms as product lines and re-using similarity-based approaches from software product line research~\citep{Henard2014}. Any more complex strategy needs to carefully consider the increase in execution times. For example, pairwise testing of the MLPClassifier of scikit-learn would require 1,101 parameter combinations, instead of 33 with our approach. Moreover, an adoption of product line techniques is not trivial: hyperparameters are often not binary and there are only relatively few parameters that can be combined, which means that the state space that needs to be explored is huge and difficult to bound.

\begin{framed}
Combinatorial smoke testing finds more bugs than smoke testing with default parameters only. 
\end{framed}

\subsection{Generalizability}

A key question is still the generalizability of our findings, i.e., the question if our results are specific for our subjects and/or the types of algorithms we considered, i.e., clustering and classification. We see strong indicators that the smoke tests would also be helpful in other contexts. 

First, we have our results, which already demonstrate that we can find bugs in three different machine learning libraries and in both supervised and unsupervised algorithms. While there are many other machine learning frameworks, we cannot think of a likely reason that other frameworks or algorithms types would behave different, unless they are, e.g., implemented in a language with stronger guarantees on numeric correctness. 

Second, there was an interesting aspect regarding the action of the maintainer with respect to bug (9) that we reported. The maintainer created a small script that ran our example that triggered the bug not only against the RidgeClassifier, but against all classifiers and regression algorithms.\footnote{\url{https://github.com/scikit-learn/scikit-learn/issues/17925\#issuecomment-658994147}} Through this, he found potential problems in regression algorithms. Hence, it is almost guaranteed that our smoke tests generalize to regression algorithms as well. 

Third, we got additional anecdotal evidence through the triage of a bug in Apache Commons Math\footnote{https://commons.apache.org/proper/commons-math/index.html} that we reported through our work on \texttt{atoml}. While the maintainer triaged our reported problem in the Kolmogorov-Smirnoff test, he added tests for values in the MAXDOUBLE range and found additional problems\footnote{https://github.com/apache/commons-math/commit/67aea22}. This is a strong indicator that at least some of our smoke tests would also be applicable for the testing of statistical tests, i.e. a type of algorithms related to machine learning. 

\begin{framed}
Evidence suggests that our results generalize to other frameworks and other types of algorithms.
\end{framed}

\subsection{Answer to our Research Question}

Based on the evidence from our case study and the discussion of the results, we can answer our underlying research question as follows. 

\begin{framed}
Tests that call core functions of machine learning libraries with extreme values as input data, especially feature values close to MAXDOUBLE, but also with all values zero, starved categories, or starved classes are well suited to provide a generic robustness check that can help to prevent severe bugs. Such tests should be applied to a wide range of hyperparameters, because otherwise bugs may be missed.

In general, it is possible to view machine learning algorithms from a textbook software testing perspective and apply standard techniques like equivalence class analysis and boundary-value analysis, if certain aspects, e.g., the interdependencies of the different types of input (e.g., features, labels, hyperparameters) are considered.  
\end{framed}

\subsection{Threats to Validity}
\label{sec:threats-to-validity}

There are several threats to the validity of our work, which we report following the classification by \cite{Wohlin2012}.

\subsubsection{Construct Validity}

The design of our case study may not be suitable to evaluate if our approach for smoke testing is really effective. Since the effectivity of tests is decided by the ability to find existing or prevent future bugs, we only used the capability to detect real-world bugs as measure for the effectiveness, instead of relying on proxies like mutation scores or test coverage. Therefore, this should not be a threat to the construct of our case study. Moreover, we reported the bugs together with the environment in which we executed them to the developers, where the bugs were independently reproduced. This removes a potential risks to our construct, because bugs were only caused by our test environment or not reproducible. Finally, we checked the issue trackers and found that non of the bugs was reported prior to our work, the developer did also not identify any duplicates. 

\subsubsection{Internal Validity}

Our conclusions regarding the effectiveness and severity are mostly based on our own assessment of the bugs we found. We mitigate this threat by describing each bug within the results section, including the likely cause of the bug, as well as the failure that we observe. This allows readers to critically assess the validity of our findings, in comparison to just reporting statistics like the number of bugs found. Moreover, we reported all bugs to the developers. Three bugs are already fixed, fixes for all but one bug would be merged if appropriate patches are provided. 

Similarly, our conclusion regarding the impact of combinatorial testing could be wrong. However, since we found multiple bugs that would be hidden without the combinatorial approach, we believe that this is a statement of fact. 

\subsubsection{External Validity}

Our claims regarding the generalizability of our findings may not be valid. One reason may be that our purposive sample is not representative for machine learning libraries. We agree that this may be the case, as many machine learning libraries are less mature. Thus, our sample represents large mature machine learning libraries, for which we believe the sample to be representative, even though it is not random~\citep{baltes2020}. However, we hypothesize that finding bugs in our sample is harder than for less mature libraries. Hence, our findings where our approach is effective should generalize to less mature libraries as well. However, we cannot rule out that other smoke tests would also be effective for smaller libraries. 

\subsubsection{Reliability}

Our work is algorithmic and we are not aware of any threats to the reliability. 

\section{Recommendations for Developers}
\label{sec:recommendation}

The results of our work have direct and actionable consequences for the day-to-day business of developers working on machine learning libraries. We formulate these consequences as two best practices, that, if implemented, should help to find existing and prevent future bugs in machine learning libraries. 

\subsection{Best Practice 1: Define Smoke Tests}

To ensure such basic functioning, developers should define and implement guidelines for smoke tests that describe
\begin{itemize}
    \item which data should be used as input for smoke tests;
    \item which functions of algorithms should be called with the data; and
    \item how different combinations of hyperparameters are considered. 
\end{itemize}
The tests should at least cover
\begin{itemize}
\item inputs with extremely large values;
\item inputs with extremely small values;
\item inputs with all features exactly zero;
\item in case of classification, empty classes and almost empty classes; and
\item if categorical data is supported, inputs with empty categories in categorical data.
\end{itemize}

Additionally, other smoke tests from Section~\ref{sec:smoketests} may also be implemented, but we have no evidence regarding the effectiveness. 

\subsection{Best Practice 2: Use Combinatorial Testing}

Tests for the basic functionality should be applied to a broad set of hyperparameters, ideally, such that equivalence class coverage is achieved. This may be achieved with a grid search over hyperparameters, but testing each equivalence class once while using default values for all other hyperparameters is also effective, scales better, and leads to fewer invalid tests. 

\section{Conclusion}
\label{sec:conclusion}

We presented an approach for combinatorial smoke testing of machine learning that is grounded in equivalence class analysis and boundary value analysis for the definition of tests. We define a set of difficult equivalence classes that specify suitable inputs for smoke testing of machine learning algorithms. Moreover, we show how the notion of equivalence class coverage in combination with the adoption of the concept of condition coverage can be used to achieve a coverage of hyperparameters for testing with linear growth. While the approach is simple, we are not aware of a similar collection of simple best practices and tests for the quality assurance of machine learning algorithms and hope that this work helps with the creation of test suites for future machine learning applications. 

Through this work, we demonstrate that textbook methods can be useful in the world of machine learning software and that they can be modified and adapted to specify effective tests for machine learning libraries. We have shown this through the application of our concepts to three mature machine learning libraries and two classes of machine learning algorithms.

Therefore, we believe that future work on testing machine learning software should not only explore completely new techniques that are tailored to machine learning and developed from scratch, but also how textbook methods, that were proven to be successful over decades of industrial use, may be used for machine learning testing. For example, future work may consider learning algorithms as state machines to allow state-based testing. 

\bibliography{./literature}

\end{document}